\documentclass[pre,onecolumn,floatfix,showpacs]{revtex4}
\usepackage{graphicx,bm,amsmath}
\begin{document}
\title{Effects of an oscillating field on magnetic domain patterns: 
Emergence of concentric-ring patterns
surrounding a strong defect} 

\author{Kazue Kudo}
\email{kudo.kazue@ocha.ac.jp}
\affiliation{Ochadai Academic Production, Division of Advanced Sciences, 
Ochanomizu University, 2-1-1 Ohtsuka, Bunkyo-ku, Tokyo 112-8610, Japan}

\date{\today}

\begin{abstract}
Oscillating fields can make domain patterns change into various types of
structures. Numerical simulations show that 
concentric-ring domain patterns
 centered at a strong defect are observed under a rapidly oscillating
field in some cases. The concentric-ring  
 pattern appears near the threshold of spatially-uniform patterns in
 high-frequency cases. 
The threshold is theoretically estimated  and the theoretical threshold
is in good agreement with numerical one in a high-frequency region.
 The theoretical analysis gives also good estimations of several
 characteristics of domain patterns for high-frequencies. 
\end{abstract}


\maketitle

\section{\label{sec:intro} Introduction}

Rapidly oscillating fields cause interesting phenomena in a wide variety
of systems. Those phenomena are often discussed in the view of
stabilization of an unstable state. One of the simplest examples is
the problem of Kapitza's inverted pendulum, which was generalized by
Landau and Lifshitz~\cite{landau}. The key idea is to separate the
dynamics into a rapidly oscillating part and a slowly varying part.
The method has been applied to various systems, e.g., 
the classical and quantum dynamics in
periodically driven systems~\cite{rahav03,rahav05}, the stabilization
of a matter-wave soliton in two-dimensional Bose-Einstein condensates
without an external trap~\cite{saito,abdullaev,liu}, and magnetic domain
patterns traveling at a slow velocity under a rapidly oscillating
field~\cite{travel}.

Domain patterns are observed in a wide variety of systems, and they show
many kinds of structures (see, for example, 
Refs.~\cite{cross,seul,muratov} and references therein).
Magnetic domain patterns are one of their good examples. While they
usually exhibit a labyrinth structure under zero field, they also show
other kinds of structures under an oscillating field. 
For instance, parallel-stripe and several kinds
of lattice structures are observed in experiments and numerical
simulations~\cite{miura,mino,tsuka}. Moreover, traveling
patterns~\cite{travel} and more interesting patterns, e.g., spirals
and concentric-ring patterns~\cite{kandaurova,mino_p}, have been observed,
depending on the strength and frequency of the field.
Spirals and concentric-ring patterns appear under a large-amplitude and
high-frequency field in experiments, and the field
range where they appear is not wide. 

In this paper, we investigate  effects of an oscillating field by
numerical simulations and theoretical analysis, focusing on the
emergence of
concentric-ring magnetic domain patterns surrounding a strong defect.  
Recently, two theoretical methods were proposed to investigate the
effects of an 
oscillating field on pattern formation in ferromagnetic thin
films~\cite{travel}. One gives the ``time-averaged model'' and the other
gives the ``phase-shifted model'': The former is derived by averaging
out rapidly oscillating terms, and the latter includes the delay of the
response to the oscillating field. In this paper, the time-averaged
model is applied to theoretical analysis, since it is
suitable for discussing ``stationary'' domain 
patterns which oscillate periodically but are unchanged in terms of a
long-time average. 
The theoretical line of the threshold for
nonuniform patterns is derived from the time-averaged model.
The theoretical threshold is consistent with the numerically simulated
one in a high-frequency region, where a concentric-ring pattern appears
around the defect.  

In fact, there are several techniques to study domain patterns under a
rapidly oscillating field theoretically. Applying a multi-time-scale
technique~\cite{michaelis,kirakosyan}, one can obtain more complex
equations in a better approximation than the time-averaged model. In
other words, the time-averaged model corresponds to the lowest orders of
multi-time-scale expansions. In this paper, the time-averaged model is
employed since it has a simple form and is efficient enough to
discuss the appearance of concentric-ring patterns in a high-frequency
region. 

The creation of a concentric-ring pattern can have different mechanisms. 
One of them is boundary conditions, and the strong defect is a kind of
boundary condition. 
The selection of a pattern depends on boundary conditions
as well as the field frequency or other parameters~\cite{dong}.  
For example, in nematic liquid crystals under a rotating magnetic field,
it is sometimes observed that the center of concentric rings nucleated 
by a dust particle moves away from it~\cite{migler}; 
Faraday experiments of viscous fluid and granular layers in
round cells show concentric-ring patterns or spiral
patterns~\cite{kiyashko,bruyn} in some cases. 
On the other hand, concentric-ring patterns can also appear
spontaneously. In fact, spiral patterns as well as 
concentric-ring patterns appear in the absence of a strong
defect under some conditions~\cite{kandaurova,tuszynski}.
However, we will not consider spontaneously created concentric-ring
patterns in this paper since those patterns are beyond the scope of this
paper.

The rest of this paper is organized as follows.
In Sec.~\ref{sec:simu}, the model of our system is introduced and
numerical results, i.e., the phase diagram
for concentric-ring patterns and profiles of the domain patterns, 
are exhibited.
In  Sec.~\ref{sec:theo}, we
discuss the threshold for 
nonuniform patterns, employing the time-averaged model. 
Moreover, several characteristics of a domain pattern estimated from the 
time-averaged model are compared with those from the numerical
simulations.  
The mechanism for the appearance of a concentric-ring 
pattern is discussed in Sec.~\ref{sec:disc}.
Conclusions are given in  Sec.~\ref{sec:conc}.

\section{\label{sec:simu} Numerical simulation}

Our model is a simple two-dimensional model (see
Refs.~\cite{travel,jagla04,kudo}, and references therein).
The Hamiltonian of the model consists of four energy terms: Uniaxial
anisotropy energy $H_{\rm ani}$, exchange interactions $H_J$, dipolar
interactions $H_{\rm di}$, and the interactions with the external field
$H_{\rm ex}$. We consider a scalar field $\phi(\bm{r})$, where
$\bm{r}=(x,y)$. 
The anisotropy energy is given by
\begin{equation}
 H_{\rm ani}=\alpha \int {\rm d}\bm{r} \lambda(\bm{r}) \left(
-\frac{\phi(\bm{r})^2}{2}+\frac{\phi(\bm{r})^4}{4}
\right),
\label{eq:Ha}
\end{equation}
where $\lambda(\bm{r})$ is employed to express the effect of defects or
the roughness of a sample.
This term implies that the values
$\phi(\bm{r})=\pm 1$ are preferable.
The positive and negative values of $\phi(\bm{r})$ 
correspond to up and down spins, respectively.
The exchange and dipolar interactions are described by
\begin{equation}
 H_J=\beta\int {\rm d}\bm{r} \frac{|\nabla\phi(\bm{r})|^2}{2}
\label{eq:Hj}
\end{equation}
and
\begin{equation}
 H_{\rm di}=\gamma\int {\rm d}\bm{r} {\rm d}\bm{r}'
 \phi(\bm{r})\phi(\bm{r}') G(\bm{r},\bm{r}'),
\label{eq:Hdi}
\end{equation}
respectively. Here, $G(\bm{r},\bm{r}')\sim |\bm{r}-\bm{r}'|^{-3}$ at
long distances.
These two terms are competing interactions:
$H_J$ implies that $\phi(\bm{r})$ tends to have the
same value as neighbors, while $H_{\rm di}$ implies
that $\phi(\bm{r})$ prefers to have the opposite sign to ones
at some distances.
The interactions with the external field is given by
\begin{equation}
 H_{\rm ex}=-h(t) \int {\rm d}\bm{r} \phi(\bm{r}).
\label{eq:Hex}
\end{equation}
Here, we consider a spatially homogeneous and rapidly oscillating field,
\begin{equation}
 h(t)=h_0\sin\omega t.
\label{eq:h}
\end{equation}
From Eqs.~(\ref{eq:Ha})--(\ref{eq:Hex}), the dynamical equation of the
model is described by
\begin{eqnarray}
 \frac{\partial \phi (\bm{r})}{\partial t}&=&
 -\frac{\delta (H_{\rm ani}+H_{J}+H_{\rm di}+{H_{\rm ex}})}
 {\delta \phi (\bm{r})} \nonumber\\
&=&  \alpha\lambda(\bm{r})  [\phi(\bm{r})-\phi(\bm{r})^3]
+\beta\nabla^2\phi(\bm{r})
-\gamma\int {\rm d}\bm{r}' \phi(\bm{r}') G(\bm{r},\bm{r}')
+h(t).
\label{eq:A-C}
\end{eqnarray}

The numerical procedures are almost the same as
those of Refs.~\cite{travel,jagla04,kudo}.
For time evolution, a semi-implicit method is employed: 
The exact solutions and the second-order
Runge-Kutta method are used for the linear and nonlinear terms,
respectively. For a better spatial resolution, a pseudo-spectral method
is applied. Namely, the time evolutions are calculated for the equation in
Fourier space corresponding to Eq.~(\ref{eq:A-C}),
\begin{equation}
 \frac{\partial \phi_{\bm{k}}}{\partial t} =\alpha 
[(\phi-\phi^3)\lambda ]_{\bm{k}}
-(\beta k^2 +\gamma G_{\bm{k}})\phi_{\bm{k}} + h(t)\delta_{\bm{k}},
\label{eq:a-1}
\end{equation}
where $[\cdot]$ denotes the convolution sum and  $G_{\bm{k}}$ is the
Fourier transform of $G({\bm{r}},0)$. Here, we define
$G(\bm{r},0)\equiv 1/|\bm{r}|^3$. Then, one has 
\begin{equation}
 G_{\bm{k}}=a_0-a_1 k,
\end{equation}
where $k=|\bm{k}|$ and 
\begin{equation}
 a_0=2\pi\int_d^{\infty} \frac{{\rm d}r}{r^2}, \quad a_1=2\pi.
\end{equation}
Here, $d$ is the cutoff length of the dipolar interactions.
In the simulations, we set $d=\pi/2$, which results in $a_0=4$.

The effect of a strong defect is incorporated in the anisotropy term,
Eq.~(\ref{eq:Ha}). Here, we put the strong defect at the center, i.e., the
origin ($\bm{r}=\bm{0}$), as follows:
\begin{equation}
 \lambda (\bm{r})= \left\{
\begin{array}{rl}
 10 & (\bm{r}=\bm{0}).\\
 1 & (\bm{r}\neq \bm{0}).
\end{array}
\right. 
\label{eq:lambda}
\end{equation}
This condition implies that the spin at the center will not flip unless
the applied field is too strong. The parameters in
Eqs.~(\ref{eq:Ha})--(\ref{eq:Hdi}) are given as
$\alpha=2$, $\beta=2$, and $\gamma=2\beta/a_1=2/\pi$.
The simulations are performed on a $128\times 128$ lattice with periodic
boundary conditions. 

\begin{figure}
\includegraphics[width=8cm]{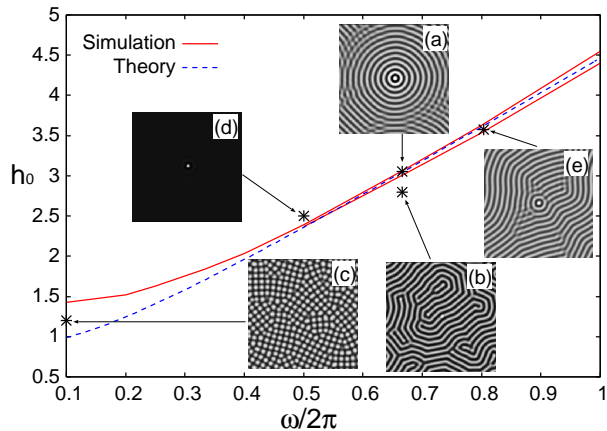}
\caption{\label{fig:diagram} (Color online) Phase diagram for the
 concentric-ring 
 patterns surrounding a strong defect at the center. The horizontal and
 vertical axes are the frequency ($\omega$) and the amplitude ($h_0$)
 of the external field, respectively. 
 The red solid lines and blue dashed line are obtained from the
 numerical simulations and theoretical analysis, respectively. 
The values of $\omega$ and $h_0$ of each snapshot are as follows: 
(a) $\omega/2\pi=2/3\simeq 0.667$ and $h_0=3.05$, 
(b) $\omega/2\pi=2/3$ and $h_0=2.8$, (c) $\omega/2\pi=0.1$ and $h_0=1.2$, 
(d) $\omega/2\pi=0.5$ and $h_0=2.5$, and (e) $\omega/2\pi=0.8$ and
 $h_0=3.56$.}    
\end{figure}

The concentric-ring patterns simulated by the numerical calculations
appear in a limited region of the frequency ($\omega$) and the amplitude
($h_0$) of the external field. Figure~\ref{fig:diagram} shows
the $\omega$-$h_0$ phase diagram for concentric-ring patterns. 
The solid lines and the dashed line are drawn by using the 
results of the numerical simulations and theoretical analysis,
respectively. The theoretical analysis is explained in
Sec.~\ref{sec:theo}. 
Concentric-ring patterns are seen only in the region between the upper and
lower solid lines.
Above the upper solid line, one sees only spatially-homogeneous patterns
except for the vicinity of the center. On the contrary, stripes,
labyrinth or lattice structures appear below the lower solid line.

Actually, it is difficult to find the exact boundaries of the region
where concentric-ring patterns appear. One may think that
concentric-ring patterns can appear right on the numerical threshold
line in a low-frequency region. 
However, even if a concentric-ring pattern could appear on the
threshold, it would be hard to find its exact value. In a
low-frequency region, the boundary between 
spatially-homogeneous patterns and nonuniform patterns (e.g., stripes,
labyrinth, and lattice structures) is sharp. In contrast, in a
high-frequency region, the boundaries between concentric-ring patterns
and other patterns are unclear: They are crossover lines rather
than transition ones.
Actually, for high frequencies, a few concentric rings around the strong
defect coexist with 
other patterns (i.e., stripes, labyrinth or lattice patterns) in the
region near the boundary lines of the concentric-ring-pattern region. 
In other words, a few concentric rings start to appear at the lower
boundary, and the number of rings grows as $h_0$ increases. At the upper
boundary, nonuniform patterns disappear except for the vicinity
of the strong defect.

\begin{figure}
\includegraphics[width=6cm]{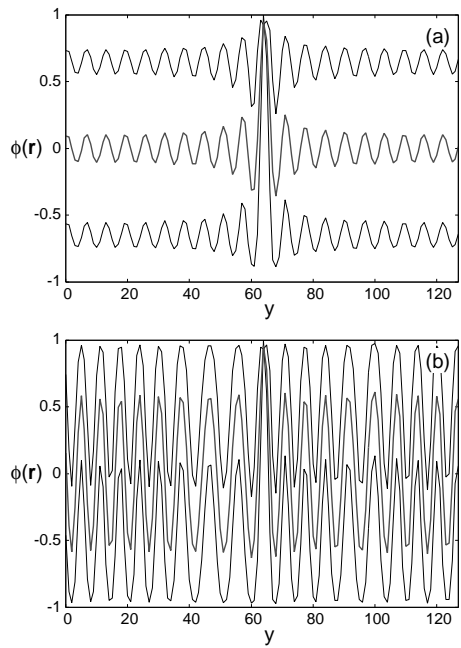}
\caption{\label{fig:prof} Profiles of the domain patterns
 corresponding to the snapshots (a) and (b) in Fig.~\ref{fig:diagram}:
 (a) $\omega/2\pi=2/3\simeq 0.667$ and $h_0=3.05$,  
(b) $\omega/2\pi=2/3$ and $h_0=2.8$. 
The upper and lower thin curves in each figure 
are the profiles at $t=(n+1/2)T$ and $t=nT$, respectively, 
where $n$ is an integer and  $T=2\pi/\omega$.
The middle thick curve is the time-averaged profile. 
}   
\end{figure}

The profile of each domain pattern is useful to see the time dependence
of the pattern. The profiles corresponding to the snapshots (a) and (b)
in Fig.~\ref{fig:diagram} are shown in Fig.~\ref{fig:prof}.
The profile is a section which includes the center (the defect) and is
perpendicular to $x$ axis (the horizontal axis). 
The profiles show that the pattern is oscillating without deformation
except for the vicinity of the center. The amplitudes of the oscillation
and that of the pattern (i.e., the difference between the maximum and
minimum values of $\phi(\bm{r})$ except for the vicinity of the defect
at a certain time) 
depend on the amplitude and frequency of the field.
 
\section{\label{sec:theo} Theoretical analysis}

Since a rapidly oscillating field is applied,
Kapitza's idea~\cite{landau} is applicable to the analysis of the
pattern formation in this system. 
In fact, the profiles in Fig.~\ref{fig:prof} validate the use of the
idea. In other words, the results in Fig.~\ref{fig:prof} justify the
fact that the variable $\phi(\bm{r},t)$ in Eq.~(\ref{eq:A-C}) consists 
of a spatially homogeneous oscillating
term $\phi_0(t)$ and a slowly varying term $\Phi(\bm{r},t)$.
In this section, the domain patterns in the absence of a
defect are discussed by employing the
time-averaged model~\cite{travel}. Namely, $\lambda(\bm{r})$ in
Eq.~(\ref{eq:lambda}) is replaced by unity, i.e., $\lambda(\bm{r})=1$.

First of all, let us consider the spatially homogeneous oscillating
solution $\phi_0(t)$ of Eq.~(\ref{eq:A-C}). Then, one has
\begin{equation}
 \dot{\phi_0}=\alpha (\phi_0-\phi_0^3)-
\gamma\phi_0\int {\rm d}\bm{r}' |\bm{r}'|^{-3} +h(t).
\label{eq:3_1}
\end{equation}
Its solution can be approximately written as 
\begin{equation}
 \phi_0=\rho\sin (\omega t +\delta),
\label{eq:form}
\end{equation}
where $\delta$ is a phase shift which comes from the delay of the
response to the field. 
Substituting Eq.~(\ref{eq:form}) into
Eq.~(\ref{eq:3_1}) and omitting high-order harmonics
(i.e., $\sin 3\omega t$), one can evaluate  $\rho$ and
$\delta$. The value of $\rho$ is obtained from the following
equations~\cite{travel}: 
\begin{equation}
 \frac{9}{16}\alpha^2 \rho^6
 -\frac32\alpha\eta_0 \rho^4 +(\omega^2+\eta_0^2)\rho^2
=h_0^2,
\label{eq:rho}
\end{equation}
where $\eta_0=\alpha-\gamma a_0$.

Now, we consider the equation for the slowly varying variable
$\Phi(\bm{r},t)$ which describes a spatially dependent pattern. 
Substituting $\phi(\bm{r},t)=\phi_0(t)+\Phi(\bm{r},t)$ into
Eq.~(\ref{eq:A-C}) and averaging out the rapid oscillation, one obtains
the time-averaged model
\begin{equation}
  \frac{\partial\Phi(\bm{r})}{\partial t}
= \left( 1-\frac32\rho^2  \right) \alpha\Phi(\bm{r})
+\beta\nabla^2\Phi(\bm{r}) -\gamma\int {\rm d}
\bm{r}' \frac{\Phi(\bm{r}')}{|\bm{r}-\bm{r}'|^3}
- \alpha\Phi(\bm{r})^3.
\label{eq:t_av}
\end{equation}

The linear stability of Eq.~(\ref{eq:t_av}) leads to the theoretical
curve in Fig.~\ref{fig:diagram} corresponding to the threshold for the
existence of nonuniform patterns.
Substituting 
$\Phi(\bm{r})=\sum_{\bm{k}}\exp(i\bm{k}\cdot\bm{r})\Phi_{\bm{k}}$
into Eq.~(\ref{eq:t_av}), one has the linear part of the equation
written as
\begin{equation}
 \frac{\partial\Phi_{\bm{k}}}{\partial t}=\eta_{\bm{k}}\Phi_{\bm{k}}.
\end{equation}
Here,
\begin{equation}
 \eta_{\bm{k}}=\left( 1-\frac32\rho^2 \right)\alpha
 -\beta (k-k_0)^2+\beta k_0^2 -\gamma a_0,
\label{eq:eta}
\end{equation}
where $k=|\bm{k}|$ and $k_0=a_1\gamma/(2\beta)$. Since  
$\eta_{\bm{k}}$ has the maximum value at $k=k_0$, the value of $\rho$
for $\eta_{k=k_0}=0$ gives the instability threshold $\rho_c$,
\begin{equation}
 \rho_c=\left[ \frac{2}{3\alpha}(\alpha+\beta k_0^2-\gamma a_0) 
  \right]^{1/2}.
\label{eq:rho_c}
\end{equation}
When $\rho > \rho_c$, $\eta_{\bm{k}}$ is negative for all values of
$\bm{k}$. In other words, the homogeneous pattern,
i.e., $\Phi(\bm{r})=0$, is stable and no inhomogeneous pattern tends to
appear for $\rho > \rho_c$.
The threshold curve for nonuniform patterns in Fig.~\ref{fig:diagram}
(the dashed line) is
given by Eq.~(\ref{eq:rho}) with $\rho=\rho_c$. 

Now, let us discuss how a nonuniform pattern disappears near the
threshold.  
Taking $\Phi(\bm{r},t)=A\cos k_0x$ which is one of the simplest stable
patterns and substituting it into Eq.(\ref{eq:t_av}), we have
\begin{equation}
 \left[
\left( 1-\frac32\rho^2 \right)\alpha +\beta k_0^2 -\gamma a_0
 -\frac34\alpha A^2
\right] A\cos k_0x -\frac{\alpha}{4}A^3\cos 3k_0x =0.
\end{equation}
Neglecting the higher harmonics (i.e., $\cos 3k_0x$), we obtain
\begin{equation}
 A=\sqrt{\frac{4}{3\alpha}}\left[
\left( 1-\frac32\rho^2 \right)\alpha +\beta k_0^2 
-\gamma a_0 \right]^{1/2}.
\label{eq:A}
\end{equation}
The amplitude $A$ of the pattern decreases monotonically in terms of
$\rho$ and vanishes at $\rho=\rho_c$. Namely, the amplitude of the pattern
diminishes near the threshold. This behavior is found in
Fig.~\ref{fig:prof}. 
The same behavior can be derived for a concentric-ring pattern, which
needs more complex calculations.

The validity of the above discussion is examined by comparing numerical
results with theoretical estimates. Actually, Fig.~\ref{fig:diagram}
indicates that the theoretical threshold is in good agreement with the
numerical one for high frequencies ($\omega/2\pi\gtrsim 0.5$).
More quantitative comparisons are given 
in Table~\ref{table} for $\omega/2\pi \ge 0.5$.
The numerical and theoretical values of $\rho$ and $A$ are compared in
it for the data near the threshold. They are obtained from the profiles
of domain patterns. Namely, the 
one-cycle time sequence of the profiles is used in order to estimate them.
The theoretical value of $\rho$ is calculated from Eq.~(\ref{eq:rho}),
and that of $A$ is calculated in two
ways: The value of $\rho$ in Eq.~(\ref{eq:A}) is given by (I) the
value from simulations and (II) the theoretical value.  
The values of $A$ estimated from simulations and Theory (I) are in
good agreement.  
Moreover, when the numerical and theoretical values of $\rho$ are close,
the value of $A$ from Theory (II) also has a similar value to 
the corresponding numerical $A$.  

\begin{table*}
\begin{center}
 \begin{tabular}{|c|c||c|c||c|c|c|}
\hline
 & & \multicolumn{2}{c||}{$\rho$}&
  \multicolumn{3}{c|}{$A$} \\
\cline{3-7}
 $\omega$ & $h_0$ & Simulation & Theory & Simulation & Theory (I) &
  Theory (II)\\
\hline
0.5 & 2.30 & $\sim$0.51 & 0.68 & $\sim$0.63 & 0.67 & 0.20\\
\cline{2-7}
 & 2.38 & 0.69 & 0.70 & 0 & 0 & 0\\
\cline{2-7}
 & 2.45 & 0.71 & 0.72 & 0 & 0 & 0\\
\hline
0.667 & 3.00 & $\sim$0.65 & 0.69 & $\sim$0.33 & 0.35 & 0.17\\
\cline{2-7}
 & 3.05 & $\sim$0.69 & 0.70 & $\sim$0.12 & 0.13 & 0\\
\cline{2-7}
 & 3.10 & 0.70 & 0.71 & 0 & 0 & 0\\
\hline
0.8 & 3.55 & $\sim$0.67 & 0.69 & $\sim$0.27 & 0.27 & 0.17\\
\cline{2-7}
 & 3.60 & $\sim$0.69 & 0.69 & $\sim$0.14 & 0.13 & 0.07\\
\cline{2-7}
 & 3.65 & 0.70 & 0.70 & 0 & 0 & 0\\
\hline
1.0 & 4.35 & $\sim$0.66 & 0.68 & $\sim$0.29 & 0.31 & 0.22\\
\cline{2-7}
 & 4.45 & $\sim$0.69 & 0.69 & 0.11 & 0.13 & 0.07\\
\cline{2-7}
 & 4.55 & 0.71 & 0.71 & 0 & 0 & 0\\
\hline
 \end{tabular}
\caption{\label{table} Numerical and theoretical values of $\rho$ and
 $A$. The numerical value of $\rho$
 is obtained from the one-cycle sequence of the
 profiles of domain patterns. The theoretical value of $\rho$ is
 calculated from Eq.~(\ref{eq:rho}), and the values of $A$ for Theory
 (I) and Theory (II) are given by Eq.~(\ref{eq:A}) with the numerical
 and theoretical values of $\rho$, respectively.} 
\end{center}
\end{table*}

\section{\label{sec:disc} Discussion}

\begin{figure}
\includegraphics[width=6cm]{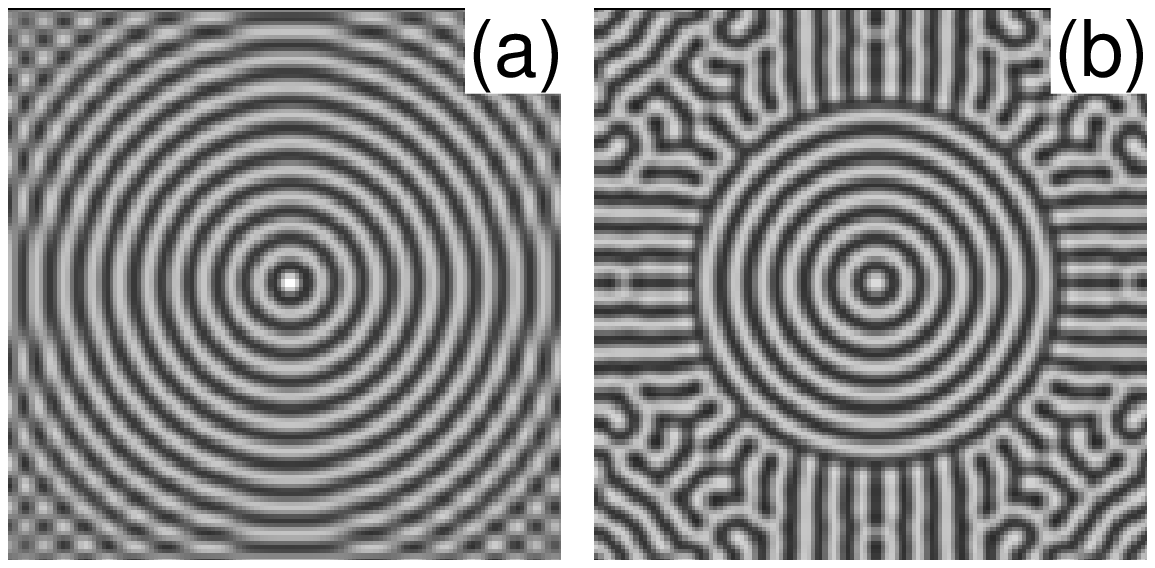}
\caption{\label{fig:snap} Domain patterns simulated by using the
 time-averaged model~(\ref{eq:t_av}) without a strong defect: 
 (a) $\rho = 0.69$, and $\rho = 0.5$. The parameters for (a) and (b) 
 correspond to those of Figs.~\ref{fig:diagram}(a) and
 \ref{fig:diagram}(b), respectively. 
}   
\end{figure}

Now, we consider how concentric-ring patterns appear around a strong
defect in this system, employing the time-averaged model.  
Actually, the time-averaged model~(\ref{eq:t_av}) without a strong
defect can also produce concentric-ring patterns. If the initial
condition is given as $\Phi(\bm{r}=0)=1$ at the center and 
$\Phi(\bm{r})=-1$ anywhere else, then concentric-ring patterns, as shown
in Fig.~\ref{fig:snap}, are demonstrated by the time-averaged model
without a strong defect. The values of $\rho$ for
Figs.~\ref{fig:snap}(a) and \ref{fig:snap}(b) are given by $\rho=0.69$ and
$\rho=5$, which correspond to those for Figs.~\ref{fig:diagram}(a)
and \ref{fig:diagram}(b), respectively.
One sees that Figs.~\ref{fig:snap}(a) and \ref{fig:diagram}(a) look very
similar. They are observed near the threshold for nonuniform
patterns, where the time-averaged model is valid.
Near the threshold, the linear growth rate $\eta_{\bm{k}}$ given by Eq. 
(\ref{eq:eta}) is very small, which means that the time scale of pattern
formation is very slow. 
In fact, it takes more than 10 times longer time to obtain
Fig.~\ref{fig:snap}(a) than Fig.~\ref{fig:snap}(b). 
If the initial pattern were completely uniform, nonuniform patterns could not
appear easily. In other words, the initial condition employed here
causes concentric rings on a uniform initial
pattern (except for the center) as one of the simplest patterns.  
This situation is very similar to
that of Fig.~\ref{fig:diagram}(a): The strong
defect in the original model behaves as a kind of boundary condition
in the almost uniform pattern near the threshold. 
Since the growth rate of the instability is very slow near the
threshold, no patterns except for concentric rings can grow for 
Figs.~\ref{fig:snap}(a) or \ref{fig:diagram}(a). 
The growth rate is expected to be relevant to the correlation length,
which is related to the number of concentric rings. 
From analogy with critical phenomena, one can estimate that the correlation
length would be proportional to $\eta_{k_0}^{-1/2}$, where 
$\eta_{k_0}=\eta_{k=k_0}$.
Therefore, the correlation length
diverges near the threshold, as $\eta_{k_0}\to 0$. 
This is the mechanism of the emergence of a concentric-ring pattern
around a strong defect and also the reason why the
boundaries of the concentric-ring-pattern region are unclear
in the phase diagram in Fig.~\ref{fig:diagram}.  

In contrast, Fig.~\ref{fig:snap}(b) has a few concentric rings, although
stripes (or mazes) spread outside the rings. The stripes
grow independently from the concentric rings because of rather large
$\eta_{\bm{k}}$'s. The difference between Figs.~\ref{fig:snap}(b) and 
\ref{fig:diagram}(b) comes out not only because the
time-averaged model is invalid far from the threshold, but also
because the initial condition employed here is pretty different from the
situation for Fig.~\ref{fig:diagram}(b).

The time-averaged model is also invalid for low frequencies.  
This is evident from Fig.~\ref{fig:diagram} in which  
the theoretical threshold line is not consistent with the numerical one
for $\omega/2\pi\lesssim 0.5$. 
The failure of the time-averaged model in a low-frequency region 
comes from the
assumption $\phi(\bm{r},t)=\phi_0+\Phi(\bm{r},t)$. 
The assumption is valid when the time scales of the rapidly oscillating
part and the slowly varying part are well separated. For low
frequencies, their separation is not sufficient.
If a multi-time-scale technique were applied,
a better theoretical threshold line would be obtained.

The concentric-ring patterns shown in this paper are purely numerical
results, although they suggest a possible mechanism about the formation
of the patterns. Since the
characteristics of domain patterns strongly depends on experimental
conditions and samples, it is rather
hard to compare experimental data with the results obtained in this
paper. However, we can suggest that concentric-ring patterns appear in a 
certain range of the field strength and frequency, which is located above
a labyrinth-pattern region. 
In fact, also in experiments, spirals and concentric rings are often
observed near the threshold of nonuniform
patterns~\cite{kandaurova,mino_p}, although they are not always centered
at a strong defect but often move around. 
Incidentally, for typical ferromagnetic
garnet films, the order of the characteristic domain width is about
10-100 $\mu{\rm m}$ and that of the frequency for lattices, spirals, or
concentric rings is about 0.1-100 kH~\cite{kandaurova,mino_p}.

\section{\label{sec:conc} Conclusions}

In this paper, effects of an oscillating field have been investigated,
and the emergence of concentric-ring patterns surrounding a strong
defect has been discussed. 
The numerical simulations show that the concentric-ring pattern
appears in the high-frequency region near the threshold for nonuniform
patterns. The simulated profiles of the concentric-ring pattern indicate
that the pattern consists of two parts (except for the vicinity of the
defect) : A rapidly 
oscillating spatially homogeneous part and a nonuniform pattern
part. This fact assures that the time-averaged model is suitable for
the theoretical analysis. The theoretical threshold line is in good
agreement with the numerical one in a high-frequency region.
When the rapidly oscillating field makes the state close
to the threshold, the concentric-ring pattern appears due to the strong
defect which is an effective boundary condition.  

In conclusion, the validity of the time-averaged model has been
demonstrated in the presence of a rapidly oscillating filed. It gives
the good estimate of the threshold for nonuniform patterns when the
field frequency is high. Moreover, it is revealed that ideal and
interesting patterns such as concentric-ring patterns can appear near the
threshold, depending on boundary conditions.

\begin{acknowledgments}
The author would like to thank M.~Mino for the information about
 experiments and K.~Nakamura for useful comments and discussion.    
\end{acknowledgments}

\end{document}